\documentclass[twocolumn,twoside,slac_two]{revtex4}
\usepackage{graphicx}
\usepackage{epsfig}
\usepackage{fancyhdr}
\begin{document}

\title{The gamma-ray spectrum from annihilation of Kaluza-Klein dark matter and its observability}

\author{Satoshi Tsuchida and Masaki Mori}
\affiliation{Department of Physical Sciences, Ritsumeikan University, Kusatsu 525-8577, Shiga, Japan}

\begin{abstract}

The lightest Kaluza-Klein particle (LKP), which appears in the theory of universal extra dimensions,
is one of the good candidates for cold dark matter.
The gamma-ray spectrum from annihilation of LKP dark matter shows a characteristic peak structure around the LKP mass.
We investigate the detectability of this peak structure by considering energy resolution of near-future detectors,
and calculate the expected count spectrum of the gamma-ray signal.
In order to judge whether the count spectrum contains the LKP signal, the $ {\chi} $ squared test is employed.
If the signal is not detected, we set some constraints on the boost factor
that is an uncertain factor dependent on the substructure of the LKP distribution in the galactic halo.
Detecting such peak structure would be conclusive evidence that dark matter is made of LKP.

\end{abstract}

\maketitle

\thispagestyle{fancy}

\section{\label{sec:intro}Introduction}

At present, most of the matter in the Universe is believed to be dark.
The existence of non-luminous matter, so-called dark matter,
was suggested by F. Zwicky in 1930s \cite{Zwicky1933}.
The dark matter problem is one of the most important mysteries in cosmology and particle physics \cite{Bertone2005}.
One feasible candidate for dark matter is the weakly interacting massive particle (WIMP),
which appears in the theory of beyond standard model.
WIMPs are good candidates for cold dark matter (CDM),
where {\it{cold}} implies a non-relativistic velocity at the decoupling time in the early Universe.
CDM comprises a large percentage of the matter density in the Universe \cite{Spergel2003},
and is necessary to form the present structure of the Universe.

The theory of universal extra dimensions (UED) is a popular theory beyond the standard model \cite{Cheng2002},
where {\it{universal}} means that all fields of the standard model can propagate into extra dimensions.
New particles predicted by this theory are called Kaluza-Klein (KK) particles.
Here, we consider the theory of UED containing only one extra dimension.
The extra dimension is compactified with radius $R$.
At tree level, the KK particle mass is given by \cite{Bergstrom2005}
\begin{eqnarray}
  \label{kaluzakleinmass}
  m^{(n)} = \sqrt{ \left( \frac{n}{R}\right)^{2} + m_{\rm{EW}}^{2}}
\end{eqnarray}
where $n$ is a mode of the KK tower, and $m_{\rm{EW}}$ is a zero mode mass of an electroweak particle.

We assume that the lightest KK particle (LKP) is
a feasible candidate for dark matter, and we denote it $B^{(1)}$.
Then, $B^{(1)}$ is the first KK mode of the hypercharge gauge boson.
Dark matter should be electrically neutral and stable particles.
Hence, LKP either does not interact with the standard model particles
or only weakly interacts with them.
In addition, LKP should have a small decay rate to survive for a cosmological time.
In the extra dimension, the four-dimensional remnant of momentum conservation is
to be conserved as KK parity, and so LKP is stable.
This hypothesis corresponds to the LKP mass $m_{B^{(1)}}$ being in the range
$0.5 \ {\rm{TeV}} \lesssim m_{B^{(1)}} \lesssim 1 \ {\rm{TeV}}$
using the above condition for CDM density \cite{Servant2003}.
In this paper, we assume the $m_{B^{(1)}}$ is 800 GeV firstly,
then we consider the change of result in the mass range of 500 GeV to 1000 GeV.

\begin{figure}[h]
  \begin{center}
    \epsfig{file=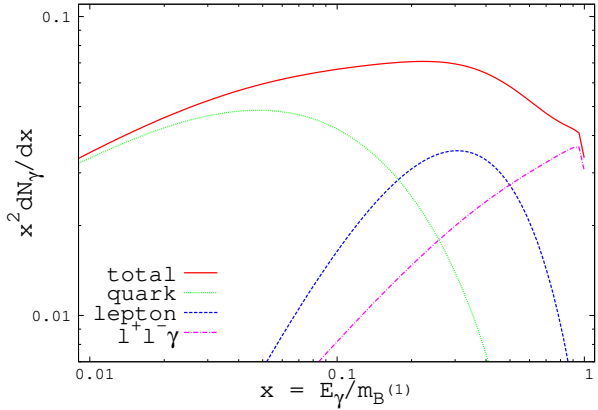,width=8cm}
    \caption{(Color online)
      Gamma-ray spectra of continuum components.
      The lines show the number of photons multiplied by $x^{2} = (E_{\gamma} / m_{B^{(1)}})^{2}$ as follows:
      the solid line shows the total number of photons per $B^{(1)} B^{(1)}$ annihilation,
      the dotted line shows the number via quark fragmentation,
      the dashed line shows the number via lepton fragmentation,
      and the dot-dashed line shows the number from the $ l^{+} l^{-} {\gamma}$ component.
      We have assumed $m_{B^{(1)}} = 800$ GeV and mass splitting is 5{\%} at the first KK level.}
    \label{fig:continuum2}
  \end{center}
\end{figure}

There are many LKP annihilation modes which contain gamma-rays as final products.
These include gamma-ray ``lines'' from two-body decays, and ``continuum'' emission.
Branching ratios into these modes can be calculated for
$B^{(1)}$ pair annihilation \cite{Cheng2002,Bergstrom2005,Servant2003}
and are not dependent on parameters other than $m_{B^{(1)}}$.
This paper considers three patterns for the continuum:
$B^{(1)}$ pairs annihilate into
(i) quark pairs,
(ii) lepton pairs which cascade or produce gamma-rays, or
(iii) two leptons and one photon ($l^{+}l^{-}{\gamma}$).
The gamma-ray spectrum of the continuum component is reproduced in Fig.\ref{fig:continuum2} as per Ref.\cite{Bergstrom2005}.
In this figure, the solid line shows the total number of photons per $B^{(1)}$ pair annihilations,
the dotted line shows the number of photons via quark fragmentation,
the dashed line shows the number of photons via lepton fragmentation,
and the dot-dashed line shows the number of photons from the $ l^{+} l^{-} {\gamma}$ mode as a function of $x=E_{\gamma}/m_{B^{(1)}}$.

When $B^{(1)}$ pairs annihilate into photon pairs, they appear as a ``line'' at the $m_{B^{(1)}}$ in the gamma-ray spectrum.
This is the most prominent signal of KK dark matter, while in most theories line models are loop-suppressed
and thus usually subdominant \cite{Bringmann2012a}.
Thus, this study focuses on the detectability of this ``line'' structure
by near-future detectors accounting for their finite energy resolution.

The distribution of dark matter is expected to be non-uniform in the Universe,
and to be concentrated in massive astronomical bodies due to gravity.
The gamma-ray flux from annihilation of dark matter particles
in the galactic halo can be written as \cite{Bergstrom2009,Aliu2009}
\begin{eqnarray}
  \label{gammarayflux}
  {\Phi}_{\gamma} &=& ({\rm{Astrophys}}) \times ({\rm{Particle \ phys}})  \nonumber \\
                  &=& J({\psi}) \times {\Phi}^{\rm{PP}}
\end{eqnarray}
where the astrophysics factor, represented by a dimensionless function $J({\psi})$, is calculated as follows:
\begin{eqnarray}
  \label{astrophys}
  J({\psi}) = \frac{1}{8.5{\rm{kpc}}} \left( \frac{1}{0.3\ {\rm{GeV \ cm^{-3}}}} \right)^{2} \ \  \nonumber \\
               \times \int_{\rm{l.o.s}} {\rho}^{2}(l) dl(\psi)
\end{eqnarray}
The function ${\rho}(l)$ is the dark matter density along the line-of-sight $l({\psi})$,
where ${\psi}$ is the angle with respect to the galactic center.
The particle physics factor is written as
\begin{eqnarray}
  \label{particlephys}
  {\Phi}^{\rm{PP}} = {\rm{Const}} \times N_{\gamma} \langle {\sigma} v \rangle
\end{eqnarray}
where $N_{\gamma}$ is the number of photons created per annihilation,
and $\langle {\sigma} v \rangle$ is the total averaged thermal cross section multiplied by the relative velocity of particles.
The value of $\langle {\sigma} v \rangle$ is accurately computed for a given dark matter candidate,
so its uncertainty is small in terms of considering the cross section containing only an s-wave.
However, this is not always the case, because some models have velocity-dependent cross sections \cite{Hamed2009,Ibe2009}.
In addition, the astrophysics factor is highly dependent on the substructure
of the dark matter distribution in the galactic halo along the line-of-sight.
Thus, we should consider the so-called ``boost factor''
which indicates the relative concentration of the dark matter in astronomical bodies compared with some benchmark distributions.
The boost factor is affected by $\langle \sigma v \rangle$ and ${\rho}^{2}(l)$, and is defined by
\begin{eqnarray}
  \label{boostfactor}
  B_{\rm{tot}} &=& B_{\rho} \times B_{{\sigma}v} \nonumber \\
               &=& \left( \frac{ \langle {\rho}^{2} (l) \rangle_{{\Delta}V}}{\langle {\rho}_{0}^{2} (l) \rangle_{{\Delta}V}} \right)
\left( \frac{\langle {\sigma}v \rangle_{v \simeq v_{\rm{disp}}}}{\langle {\sigma}v \rangle_{v \simeq v_{\rm{F}}}} \right)_{{\Delta}V}
\end{eqnarray}
where $v_{\rm{disp}}$ is the velocity dispersion, $v_{\rm{F}}$ is the typical velocity at freeze-out, 
a volume ${\Delta}V$ is a diffusion scale, and ${\rho}_{0}(l)$ is a typical dark matter density profile, such as Navarro-Frenk-White (NFW) \cite{Navarro1996}.
$B_{\rho}$ could be as high as 1000 when accounting for the expected effects of adiabatic compression \cite{Prada2004}.

In the case of gamma-ray flux from LKP annihilation,
the particle physics factor is almost fixed for a given model, and
the boost factor mostly depends on the astrophysical contribution.
In this paper, we vary only $B_{\rm{tot}}$ as a parameter
which describes our limited knowledge regarding the astrophysical contribution,
and consider constraints on its value from observation.

Recent progress in gamma-ray observation has revealed new findings in the galactic center region.
The high energy gamma-rays from the galactic center have been observed by
the High Energy Stereoscopic System (HESS) \cite{Aharonian2009}, the Large Area Telescope 
on board the Fermi Gamma-Ray Space Telescope (Fermi-LAT) \cite{Nolan2012} and other experiments.
However, the observed gamma-ray spectrum is represented as a power-law plus an exponential cut-off,
and is hardly compatible with a dark matter signal \cite{Aharonian2006}.
Recently, some evidence regarding the enhancement of the continuum component of the gamma-ray emission
from the galactic center region 
detected by the Fermi-LAT has been reported and argued as a possible
signal from the decay of dark matter particles \cite{Hooper2011,Hooper2011a,Abazajian2012}.
Further, an enhancement of around 130 GeV in the energy
spectrum of gamma-rays from the galactic center region
has been reported which may indicate a possible dark matter signal
\cite{Bringmann2012b,Weniger2012,Su2012,Finkbeiner2013}.
Discussion relating to unifying the continuum and the line has also been presented \cite{Asano2013}.
However, the analysis by the Fermi-LAT collaboration did not
confirm the significance of the line detection \cite{Ackermann2012,Ackermann2013}.
Thus, the situation is still unclear and more sensitive observation is necessary to resolve the issue.

In the following, we focus on gamma-ray observation with near-future missions, such as the Calorimetric Electron Telescope (CALET) \cite{Torii2011}.
CALET is a fine resolution calorimeter for cosmic-ray observation to be installed on the International Space Station.
CALET will detect gamma-rays in the energy range of 4 GeV to 1 TeV with
about 1000 ${\rm{cm}^{2}}$ effective area, and
a few percent energy resolution, suitable for gamma-ray line detection \cite{Mori2013}.

In this paper, we analyze the gamma-ray spectra from $B^{(1)}$ pair annihilation
accounting for the finite energy resolution of gamma-ray detectors
and purposefully discuss the observability of the ``line'' at the $m_{B^{(1)}}$.
We then give possible constraints on the boost factor by near-future detectors.

\section{The effect of energy resolution}

The gamma-ray spectrum ${d{\Phi}_{\gamma} ({\Delta}{\Omega})}/{dE_{\gamma}}$ reaching a detector can be expressed as \cite{Bergstrom2005}
\begin{eqnarray}
  \label{relationxande}
  E_{\gamma}^{2} \frac{d{\Phi}_{\gamma}({\Delta}{\Omega})}{dE_{\gamma}} \simeq {\rm{Const}} \times B_{\rm{tot}} \times x^{2} \frac{dN_{\gamma}}{dx},
\end{eqnarray}
where ${\Delta} {\Omega}$ is the angular acceptance of the detector,
\begin{eqnarray}
  \label{constfactor}
  {\rm{Const}} \simeq 3.5 \times 10^{-8} \left( \frac{\langle {\sigma} v \rangle}{3 \times 10^{-26} {\rm{cm^{3}s^{-1}}}} \right) \ \ \  \nonumber \\
     \times \left( \frac{0.8 {\rm{TeV}}}{m_{B^{(1)}}} \right) \langle J_{\rm{GC}} {\rangle}_{{\Delta}{\Omega}} {\Delta}{\Omega},
\end{eqnarray}
and $\langle J_{\rm{GC}} {\rangle}_{{\Delta}{\Omega}}$ is a dimensionless line-of-sight integral averaged over ${\Delta}{\Omega}$.
If we assume an NFW profile,
$\langle J_{\rm{GC}} {\rangle}_{{\Delta}{\Omega}} {\Delta}{\Omega}$ equals to 0.13 for a ${\Delta}{\Omega}= 10^{-5}$ \cite{Cesarini2004}.
In this case $dN_{\gamma}/dx$ includes both the continuum and line components.

\begin{figure}[htbp]
  \begin{center}
    \epsfig{file=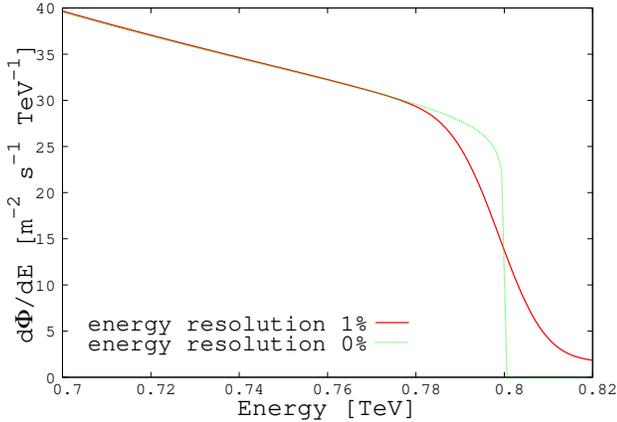,width=8cm}
    \caption{(Color online)
      Gamma-ray spectra of the continuum accounting for energy resolution assuming $m_{B^{(1)}} = 800$ GeV.
      The solid line assumes an energy resolution of 1{\%} with a Gaussian distribution,
      and the dotted line does not include the effect of energy resolution,
      as per the solid line in Fig.\ref{fig:continuum2}.
      The assumed boost factor is 1000.}
    \label{fig:continuumenereso2}
  \end{center}
\end{figure}

Now, we discuss the effect of energy resolution.
If the measured energies of detected gamma-rays behave like a Gaussian distribution and
the energy resolution is 1{\%},
the measured ``continuum'' gamma-ray spectrum is blurred and should appear as shown in Fig.\ref{fig:continuumenereso2}.
Here we draw the curve assuming the following equation
\begin{eqnarray}
  \label{energygaussian}
  g(E) \propto \int f(E') \times {\rm{exp}} \left[ - \frac{(E-E')^{2}}{2{\sigma}_{E}^{2}} \right] dE' ,
\end{eqnarray}
where $f(E')$ corresponds to a function shown by the solid line in Fig.\ref{fig:continuum2},
and ${\sigma}_{E}$ is the energy resolution.

\begin{figure}[t]
  \begin{center}
    \epsfig{file=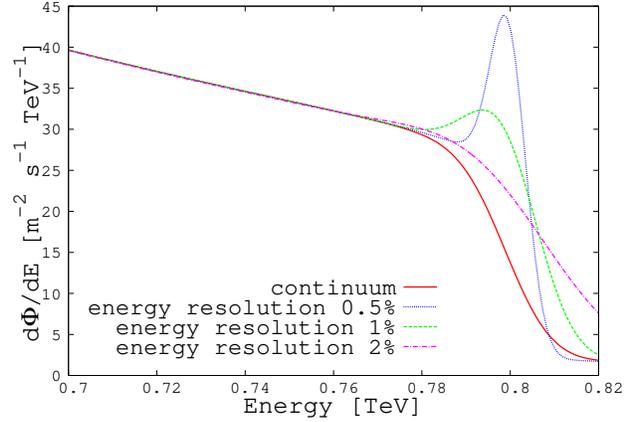,width=8cm}
    \caption{(Color online)
      Gamma-ray spectra of continuum plus line diffused by the energy resolution assuming $m_{B^{(1)}} = 800$ GeV.
      The solid line shows the continuum component only, assuming the energy resolution of 1{\%},
      while the dotted, dashed and dot-dashed lines show the continuum plus line components assuming
      energy resolution values of 0.5{\%}, 1{\%}, 2{\%} respectively.
      The assumed boost factor is 1000.}
    \label{fig:con1perwab}
  \end{center}
\end{figure}

Next, we analyze how the ``line'' from the $B^{(1)}$ pair annihilation into photon pairs looks above the ``continuum''.
The three patterned lines shown in Fig.\ref{fig:con1perwab} assume different energy resolutions
which we take as 0.5{\%}, 1{\%} and 2{\%} with the Gaussian distribution.
In Fig.\ref{fig:con1perwab}, the solid line shows the continuum component only with an energy resolution of 1{\%},
and the patterned lines show ``line'' plus ``continuum'' spectra for different energy resolutions:
the dotted line, dashed line and dot-dashed line show the spectra
when the energy resolution is 0.5{\%}, 1{\%}, 2{\%} respectively,
assuming the boost factor $B_{\rm{tot}} = 1000$.

\begin{figure}[htbp]
  \begin{center}
    \epsfig{file=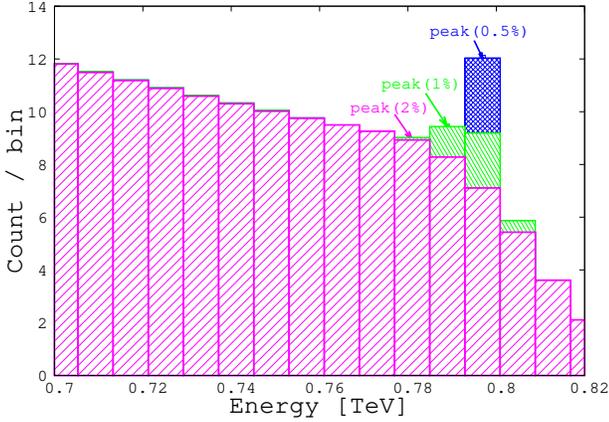,width=8cm}
    \caption{(Color online)
      Expected count spectra near the peak assuming energy resolutions of 0.5{\%}, 1{\%} and 2{\%} assuming $m_{B^{(1)}} = 800$ GeV.
      The bin width of histograms is 8 GeV, equaling 1{\%} of the $m_{B^{(1)}}$.
      The assumed boost factor is 1000.}
    \label{fig:countbins}
  \end{center}
\end{figure}

We can transform the spectra into counts to be observed by gamma-ray detectors.
This is accomplished through multiplying by a factor of 0.03
for an assumed observation time of 1 yr and an assumed effective area of 1000 ${\rm{cm}^{2}}$.
These values arise from the typical aforementioned CALET sensitivity \cite{Mori2013}.
When analyzing observational data, the energy bin width must be specified.
A bin width of 1{\%} of $m_{B^{(1)}}$ (about one standard deviation of energy reconstruction) was used in order to avoid energy information loss.
The resulting histograms are shown in Fig.\ref{fig:countbins},
where plots of the three cases corresponding to energy resolutions of 0.5{\%}, 1{\%} and 2{\%} are shown.
The figure shows that if the energy resolution of the detectors
becomes 2{\%} or worse, the characteristic peak indicating the $m_{B^{(1)}}$ will be diffused,
making it hard to resolve into the line and continuum components.
Thus, the energy resolution for gamma-ray detectors should be better than 2{\%},
in order to ``resolve the line'' without the need for detailed analysis.

\begin{figure}[htbp]
  \begin{center}
    \epsfig{file=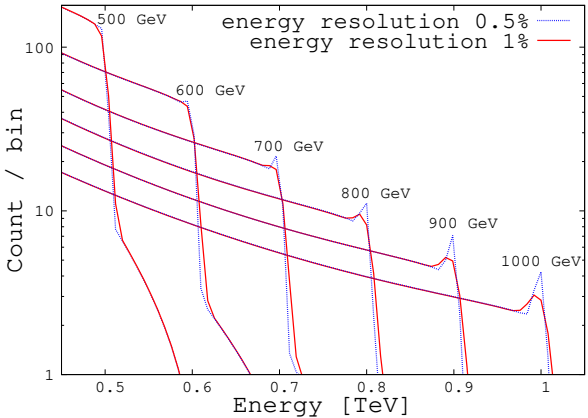,width=8cm}
    \caption{(Color online)
      Expected count spectra, assuming energy resolutions of 0.5{\%} and 1{\%}.
      The data space is 8 GeV, which is 1{\%} of for the $m_{B^{(1)}}$ = 800 GeV.
      The assumed boost factor is 1000.}
    \label{fig:goukeicomparesmooth}
  \end{center}
\end{figure}

Thus far, we have taken the LKP mass to be $m_{B^{(1)}} = 800$ GeV, and calculated count spectrum for its mass.
Now, we vary the mass from 500 GeV to 1000 GeV in 100 GeV intervals,
and calculate the count spectrum for each mass.
The results are shown in Fig.\ref{fig:goukeicomparesmooth}.
This figure shows that the characteristic peak structure is visually clearer when $m_{B^{(1)}}$ is heavier.
That is, the line component becomes relatively larger since the continuum component decreases for heavier $m_{B^{(1)}}$.

\section{Discussion}

We now discuss the observability of the LKP signal in near-future detectors,
taking account of the observed background spectrum.
That is, we give estimates for the accessible range of the boost factor
when the observed counts are significantly different from the background spectrum.
Here, we assume the gamma-ray spectrum from HESS J1745-290 located near the center of the Galaxy is the source of the background.
Its spectrum is given by \cite{Aharonian2009}
\begin{eqnarray}
  \label{background}
  \frac{d{\Phi}}{dE} = (2.55 \pm 0.06 \pm 0.40) \left( \frac{E}{\rm{TeV}}\right)^{-2.10 \pm 0.04 \pm 0.10} \  \nonumber \\
         \times \ {\rm{exp}} \left[ - \frac{E}{(15.7 \pm 3.4 \pm 2.5){\rm{TeV}}} \right] \ \ \ \ \ \  \nonumber \\
\times 10^{-8} \ \ \ {\rm{TeV}^{-1}} \ {\rm{m}^{-2}} \ {\rm{s}^{-1}}.\ \ \ \ 
\end{eqnarray}
Note that with the energy resolution of HESS (15{\%}), the LKP ``line'' signal is broadened and hard to detect.

To investigate the detectability quantitatively,
we employ a ${\chi}$-squared test method
to judge whether the excess counts are statistically meaningful.

\begin{figure}[t]
  \begin{center}
    \epsfig{file=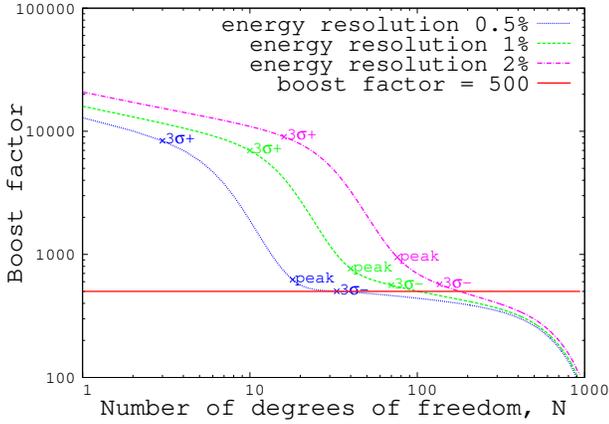,width=8cm}
    \caption{(Color online)
      Expected limits on the boost factor for the $m_{B^{(1)}} = 800$ GeV
      as a function of the number of degrees of freedom of the observed energy range.
      The horizontal solid line shows a boost factor of 500.
      The dotted, dashed and dot-dashed lines show the boost factor
      when ${\chi}^{2}$ values are bigger than critical values for energy resolution 0.5{\%}, 1.0{\%}, and 2{\%} respectively.}
    \label{fig:chisqboost}
  \end{center}
\end{figure}

First, we define ${\chi}^{2}$ as
\begin{eqnarray}
  \label{chisquare}
  {\chi}^{2} = \sum^{N} \frac{([{\rm{count}}+{\rm{background}}]-{\rm{background}})^{2}}{{\rm{background}}} \ \ 
\end{eqnarray}
where $N$ is the number of energy bins, corresponding to degrees of freedom for the ${\chi}$-squared test.
We then specify the energy range:
\begin{eqnarray}
  \label{energyrange}
  {\rm{Energy \ range}} = [ 100 \ {\rm{GeV}} , 1 \ {\rm{TeV}}] 
\end{eqnarray}
with bin width of 0.8 GeV (= 0.1{\%} for $m_{B^{(1)}} = 800$ GeV).
Thus, $N$ is about 1000 in this case.
The upper bound of the energy range under analysis is fixed as
$m_{B^{(1)}} + 3{\sigma}_{E}$ to allow finite energy resolution.
Hence, at this energy, the degree of freedom is one ($N=1$).
Then, we vary the lower bound of the energy range to lower energies.
Thus, $N$ gradually increases as we expand the energy range to lower energies.
For example, $N$ at the peak for 1{\%} energy resolution is
\begin{eqnarray}
  \label{nrange}
  N \left[ E_{\rm{peak}} , m_{B^{(1)}} + 3 {\sigma}_{E} \right] = 40.
\end{eqnarray}
We investigate the value of boost factor when ${\chi}^{2}$ is bigger than some critical value for
each $N$.
The relation between $N$ and the upper bound of the boost factor is shown in Fig.\ref{fig:chisqboost},
where the ``peak'' on each line corresponds to the value
when $N$ equals to Eq.(\ref{nrange}).
Then, $3{\sigma}^{\pm}$ are the energy width limits within $3{\sigma}$ from the peak.
Thus, they are given as
\begin{eqnarray}
  \label{sigmapm}
  N \ {\rm{at}} \ 3{\sigma}^{\pm} = N \left[ E_{\rm{peak}} \pm 3{\sigma}_{E} , m_{B^{(1)}} + 3 {\sigma}_{E} \right]
\end{eqnarray}
One can see from this figure that the limit on the boost factor does not change rapidly
when we include energy bins well below the peak.
An accessible boost factor would be smaller than 500 when $N$ is in the range 30 - 200.
These values of $N$ correspond to being near the peak energy for annihilation of LKP.

\begin{figure}[htbp]
  \begin{center}
    \epsfig{file=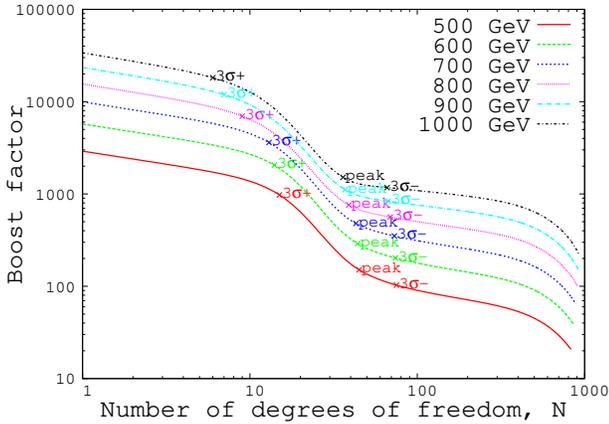,width=8cm}
    \caption{(Color online)
      Expected limits on the boost factor for each mass, assuming an energy resolution of 1{\%},
      as a function of the number of degrees of freedom of the observed energy range.}
    \label{fig:chisqboost1}
  \end{center}
\end{figure}

\begin{figure}[htbp]
  \begin{center}
    \epsfig{file=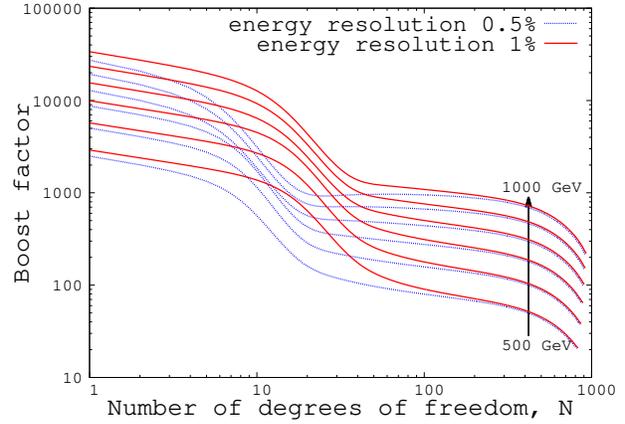,width=8cm}
    \caption{(Color online)
      Comparison of the expected limits of boost factors with 1{\%} and 0.5{\%} energy resolution.
      The individual lines refer to the scenarios with masses set from 500 GeV to 1000 GeV in 100 GeV intervals.}
    \label{fig:chisqboostcompare}
  \end{center}
\end{figure}

We applied similar analyses for other LKP masses.
The results shown in Fig.\ref{fig:chisqboost1}
indicate that the constraint for the boost factor is tighter for lighter $m_{B^{(1)}}$.
In addition, we compared the results when using energy resolution of 1{\%} and 0.5{\%},
as shown in Fig.\ref{fig:chisqboostcompare}.
The number of events near the peak increases with better energy resolution,
and the resulting constraint near the peak is tighter.

Some constraints from observations on the KK dark matter models have been reported.
The Fermi-LAT team searched for gamma-ray emission from dwarf spheroidal galaxies
around the Milky Way galaxy and set constraints on dark matter models with non-detection results \cite{Abdo2010}. 
The HESS array of imaging air Cherenkov telescopes observed the Sagittarius
dwarf spheroidal galaxy in the sub-TeV energy region and derived lower limits on the $m_{B^{(1)}}$ of 500~GeV \cite{Aharonian2008}.
These results put constraints on $\langle \sigma v\rangle$ of dark matter halo KK particles.
The present limits allow the maximum value of boost factors of several to $1.5\times 10^4$ depending on $m_{B^{(1)}}$.
Our analysis on future high energy resolution observation improves the
limits on the boost factor or the chance to detect the signal.
If such signals are detected, we will be able to say
that dark matter is made of LKP,
which will be evidence of the existence of extra dimensions.

\section{Conclusion}

Energy resolution plays a key role in detecting the line structure of the gamma-ray spectrum
expected from annihilation of LKP dark matter as predicted by UED theories.
This paper investigated the effects of energy resolution of gamma-ray detectors and calculated the expected count spectrum.
The predicted gamma-ray spectrum is the sum of the continuum and a line corresponding to $m_{B^{(1)}}$,
but this characteristic spectrum is diluted when we account for the finite energy resolution of detectors
as shown in Fig.\ref{fig:continuumenereso2} and Fig.\ref{fig:con1perwab}.
Further, if we assume the exposure (area multiplied by observation time) of near-future detectors,
count statistics will be the final limiting factor.
The characteristic peak indicating the $m_{B^{(1)}}$ would be diffused if the energy resolution is 2{\%} or worse.
However, with qualitative statistical analysis, we may be able to detect a peak
statistically by subtracting a background from the observed spectrum.
In addition, if $m_{B^{(1)}}$ is heavy, the observed gamma-ray spectrum will show the characteristic peak clearly
because the continuum component decreases relative to the line component.

This paper also estimated the accessible range of the boost factor using a ${\chi}$-squared test
assuming the HESS J1745-290 spectrum as a background.
If the observed energy range for gamma-rays extends to lower energies,
the accessible range of the boost factor will be lowered
since a higher amount of continuum events will be detected.
If the signal is not detected, the upper limit of the boost factor is about 500
if only taking data near the peak, and about 100 if the whole energy range is covered.
Furthermore, if $m_{B^{(1)}}$ is light or the energy resolution of the detector is good (say the order of 0.5{\%}),
we may tightly constrain the boost factor.

If the gamma-ray line structure is observed in the future,
we may identify LKP dark matter,
which will provide strong evidence for the existence of extra dimensions.

\section*{Acknowledgments}

We would like to thank Y. Sugawara, and K. Sakai for useful discussions.
We also thank F. Matsui for helpful comments.

\end{document}